\documentclass[twocolumn,showpacs,preprintnumbers,amsmath,amssymb]{revtex4}

\usepackage{graphicx}
\usepackage{dcolumn}
\usepackage{bm}

\begin{document}

\preprint{APS/123-QED}

\title{Dynamics of the electron-nuclear and heteronuclear polarization transfers
in optically-oriented semi-insulating InP:Fe}
\author{Atsushi Goto$^{1,2}$}
\email{goto.atsushi@nims.go.jp}
\author{Kenjiro Hashi$^{1}$}
\author{Tadashi Shimizu$^{1}$}
\author{Shinobu Ohki$^{1}$}
\affiliation{$^{1}$National Institute for Materials Science, 
Tsukuba, Ibaraki 305-0003, Japan}
\affiliation{$^{2}$PRESTO, Japan Science and Technology Agency, 4-1-8 Honcho, Kawaguchi, Saitama 332-0012, Japan.}%

\date{\today}

\begin{abstract}
Dynamics of the electron-nuclear and heteronuclear polarization-transfer processes 
in the optically oriented semi-insulating iron-doped indium phosphide were investigated
through the characteristic time scales of the processes.
(1) We find in the optical nuclear orientation process 
that the buildup times for $^{31}$P and $^{115}$In nuclear polarizations 
are of the same order.
From the analysis of the cross-relaxation process between photo-excited electrons and nuclei,
it is concluded that electron-nuclear dipolar-couplings are responsible 
for the polarization transfer in this case.
This example shows that the nuclear-site dependence of the buildup time 
can be utilized to identify hyperfine couplings responsible for the process.
(2) We find 
in the heteronuclear polarization transfer (cross polarization) process 
between optically oriented $^{31}$P and $^{115}$In
that the cross-relaxation time is rather short;
it is an order of magnitude shorter than that expected for nuclear dipolar couplings
when a magnetic field is applied parallel to the crystalline [100] axis.
From the cross polarization spectral density analysis,
it is concluded that a large $J$-coupling of the order of 2 kHz exists between these nuclei
and that its angular dependence is not of a simple pseudo-dipolar type.
\end{abstract}

\pacs{78.30.Fs  32.80.Bx  76.70.Fz}
\maketitle

\section{Introduction\label{sec:introduction}}
Dynamic nuclear polarization (DNP) is a method of creating hyperpolarized nuclear spins
in solids, liquids or gases in a non-equilibrium fashion. 
For decades, it has been utilized to enhance sensitivity 
in the nuclear magnetic resonance (NMR) method.
Recently, nuclear-spin-related phenomena in semiconductors have attracted much attention, 
which has added renewed interests to DNP.
Examples include electron spin dynamics in semiconducting nanostructures\cite{schliemann03}
and solid-state NMR quantum computers;\cite{shimizu02}
the hyperpolarized nuclei are utilized
as a means to control electron spin states in the former,
while they are expected to serve as initialized states for quantum computation in the latter.
With the emergence of these new applications,
it has become increasingly important to control
positions, degrees of polarization and the nuclear species to be polarized, 
i.e., to manipulate hyperpolarization.

To achieve this purpose,
one needs to create, transfer and localize nuclear spin polarizations
in desired positions efficiently,
which may require a variety of techniques to be integrated.\cite{tycko98a}
Nuclear spin polarizer (NSP) is one of such schemes.\cite{goto03a,goto04a}
In this scheme, hyperpolarization is created in compound semiconductors
such as InP and GaAs
by the optical orientation (optical pumping) method,\cite{meier84,tycko96}
where nuclear spins are hyperpolarized by polarized electrons 
photo-excited by circularly polarized light
with the band gap energies of the semiconductors.
The hyperpolarization thus created is manipulated 
by means of various polarization transfer techniques in solids,
such as the cross polarization, polarization transfer and spin diffusion,
which enable us to transfer hyperpolarization
between different nuclear species, different portions in the semiconductor, 
or even different materials through the interfaces.
In the last case, the polarization can be localized at the interfaces.
\cite{tycko98a,goto03a,goehring03,goto07a}

The polarization transfers in this scheme rely on hyperfine- 
and heteronuclear-couplings in semiconductors
so that it is essential to understand their characteristics.
Our understanding on them is still insufficient, however.
The characteristics of the hyperfine couplings 
responsible for the polarization transfer 
from photo-excited electrons to nuclei in the optical orientation process 
are still open to debate.
\cite{patel99,paravastu04,paravastu05}
On the other hand, 
the strengths and angular dependences of the heteronuclear couplings in InP
have been measured in the thermal equilibrium,\cite{tomaselli98,iijima03}
but those between optically oriented nuclei still remain undetermined.

In this paper, we address these issues
from the viewpoint of the characteristic time scales
of the polarization transfer processes
i.e., the cross-relaxation time between hyperpolarized nuclei, $T_{IS}$
and the buildup time for nuclear polarization 
by photo-excited electrons, $T_\textnormal{b}$.
The former ($T_{IS}$) provides us with information 
on heteronuclear couplings between hyperpolarized nuclei.
In \S \ref{cross relaxation}, 
we show in the case of the semi-insulating iron-doped InP (InP:Fe)
that the polarization transfer is predominantly mediated 
by nearest-neighbor indirect $J$-couplings,
and that their angular dependence is not of a simple pseudo-dipolar type.
The latter ($T_\textnormal{b}$), on the other hand, 
provides us with information 
on the hyperfine couplings responsible for the nuclear spin orientation,
which are closely related to the states of the photo-excited electrons.
In \S \ref{buildup time}, 
we show that the nuclear-site dependence of $T_\textnormal{b}$ 
is a good measure to determine the types of hyperfine couplings
and that, in the case of InP:Fe, 
the polarization transfer from photo-excited electrons
to nuclei is primarily brought about by the electron-nuclear dipolar couplings.

\section{Experimental Methods}

The optical-nuclear-orientation and cross-polarization experiments were performed at 10 K
with the optical pumping double resonance system.\cite{goto06}
The system includes a two-channel (XY) NMR spectrometer (Apollo, Tecmag Inc.),
a Ti:Sapphire tunable laser (3900S) pumped by a diode-pumped Nd:YVO$_4$ 
cw green laser (Millennia Vs, Spectra-Physics Inc.),
and a home-built top-loading XY double-resonance probe with an optical fiber attachment.
The probe is installed in a dynamic gas-flow cryostat (Spectrostat 86/62, Oxford Instruments Inc.),
which is mounted on a 270 MHz (6.346 T) wide-bore 
superconducting magnet (Oxford Instruments Inc.).
The Ti:Sapphire laser provides 
linearly polarized light with the wavelength ranging between 600 and 1000 nm, 
which is transmitted to the sample space at the probe end 
by a polarization maintaining optical fiber (PANDA, Fujikura Co. Ltd.),
then converted to circularly polarized light with a quarter waveplate
before being applied to a sample.
The sample used in this study was a wafer of the semi-insulating iron-doped InP
with the crystal orientation of (100) 
and the carrier density at room temperature of $3 \times 10^7$ cm$^{-3}$
(Showa Denko, lot $\sharp$20044202).
It was set inside a pickup coil at the probe end 
with the surface normal to the magnetic field and the light beam.

\begin{figure}[b]
 \begin{center}
  \includegraphics[scale=0.45]{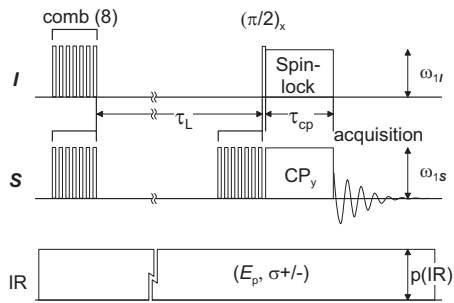}
 \end{center}
\caption{\label{sequence}
Pulse sequence for the cross polarization experiments 
under the infrared light irradiation.
$I$ and $S$ correspond to $^{31}$P and $^{115}$In, respectively,
and IR represents the infrared light with the photon energy of $E_p$
and the helicity of $\sigma^{\pm}$. 
For other notations, refer to the text.}
\end{figure}

The pulse sequence for $I$ (primary nuclei, $^{31}$P), 
$S$ (secondary nuclei, $^{115}$In)
and IR (infrared light with the photon energy of $E_p$ 
and the helicity of $\sigma^{\pm}$) 
used in the optical-orientation-cross-polarization experiments 
is schematically shown in Fig. \ref{sequence}.
It consists of the following four processes, i.e.,
(1) saturation, (2) optical pumping, (3) cross polarization and (4) detection.
(1) At the beginning, saturation pulses consisting of eight $\pi/2$ pulses
are applied to both the nuclei,
which extinguish the initial polarizations at thermal equilibrium. 
(2) The sample is irradiated only with the infrared light 
for the duration of $\tau_{\rm L}$,
which creates polarizations of both the nuclei 
inside the illuminated region of the sample.
The polarization in the bulk (outside of the illuminated region) 
can also grow toward the equilibrium state for this duration.
The $I$-polarization, however, does not recover
because the spin-lattice relaxation time $T_1$ at $^{31}$P 
is much longer than $\tau_{\rm L}$.\cite{goto04b}.
The $S$-polarization, on the other hand, is extinguished again
by the saturation pulses at the end of the duration.
Consequently, only the optically oriented $I$-polarization
in the illuminated region remains at the end of the duration.
(3) The cross-polarization is applied between the $I$- and $S$- spins,
which transfers the $I$-polarization to $S$ in the illuminated region,
but not in the bulk because of the lack of $I$-polarization there.
(4) The $S$ signal from only the illuminated region 
is detected as a free induction decay.
In our experiments, 
the effective duration time $\tau_L$ was fixed at 120 s.

\section{Cross-relaxation in hyperpolarized nuclear spins
\label{cross relaxation}}

\begin{figure}[b]
 \begin{center}
  \includegraphics[scale=1.0]{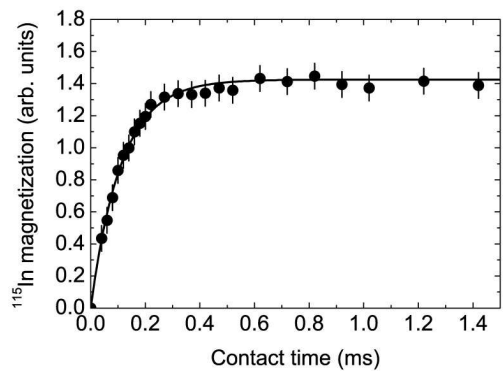}
 \end{center}
\caption{\label{contact time dependence}
The $\tau_\textnormal{cp}$ dependence of the $^{115}$In magnetization 
in the $^{31}$P $\rightarrow$ $^{115}$In cross polarization experiment
under the IR irradiation ($E_p$=1.386 eV) measured at 10 K and 6.346 T.
$\omega_{0I}/2\pi$=109.316 MHz, $\omega_{0S}/2\pi$= 59.23 MHz and
$\omega_{1I}/2\pi=\omega_{1S}/2\pi$= 18 kHz.
The solid line is a result of the least-squares-fitting to Eq. (\ref{MS-single}).}
\end{figure}

In general, 
the contact time ($\tau_\textnormal{cp}$) dependence of the $S$-spin magnetization 
in the cross polarization process is given by,
\begin{equation}
 M_S(\tau_\textnormal{cp}) \propto [ 1-\exp\{
-(1-\frac{T_{IS}}{T_{1\rho}^{I}})\frac{\tau_\textnormal{cp}}{T_{IS}}\} ]
\exp(-\frac{\tau_\textnormal{cp}}{T_{1\rho}^{I}}),
\label{MS}
\end{equation}
which results from the competition between the polarization transfer process 
from the $I$-spins with the characteristic time $T_{IS}$, 
and the decay process of the $I$-spins in the rotating flame 
characterized by $T_{1\rho}^{I}$.

The $\tau_\textnormal{cp}$ dependence of the $^{115}$In magnetization 
under the IR irradiation is shown in Fig. \ref{contact time dependence}.
One may find that the decay process is negligible
($T_{1\rho}^{I} \rightarrow \infty$).
In fact, $T_{1\rho}$($^{31}$P) was reported to be as long as 80 ms,\cite{tomaselli98}
which is much longer than $T_{IS}$.
Setting $T_{1\rho}^{I} \rightarrow \infty$,
Eq. (\ref{MS}) can be reduced to,
\begin{equation}
  M_S(\tau_\textnormal{cp}) = M_S(\infty) \{1 - \exp ( -\tau_\textnormal{cp}/T_{IS})\}.
\label{MS-single}
\end{equation}
By fitting the data in Fig. \ref{contact time dependence} to Eq. (\ref{MS-single}),
one can obtain the cross relaxation rate $T_{IS}^{-1}= (8.8 \pm 0.8) \times 10^{3}$ s$^{-1}$.


It is intriguing to see 
whether or not $T_{IS}^{-1}$ is affected 
by the difference in the photon energy $E_p$.
We measured $T_{IS}^{-1}$ at two photon energies, $E_p$ = 1.386 and 1.407 eV,
at both of which the $^{31}$P polarization is strongly enhanced 
by the optical orientation effect.
The result is summarized in Table \ref{table:1},
which shows that $T_{IS}^{-1}$ is independent of $E_p$ within the experimental error.
One of the possible explanations for this result may be 
relaxation of the photo-excited electrons,
which occurs with a time scale much faster than $T_{IS}$, 
so that the electrons excited with different $E_p$ 
would result in the same meta-stable state.
We also measured $T_{IS}^{-1}$ 
at two different IR-irradiation times $\tau_L$ = 60 s and 240 s,
which is intended to investigate the effect of nuclear spin diffusion process.
It is expected that for greater $\tau_L$,
nuclear polarizations may spread out farther from the positions
where polarizations are originally created.
Provided the spin diffusion constant $D \approx 10^{-13}$cm$^2$/sec,\cite{abragam61} 
the expected diffusion lengths $r_D \approx \sqrt{D\tau_L}$ = 24 and 49 nm
for $\tau_L$ = 60 s and 240 s, respectively.\cite{goto04a}
The result is shown in Table \ref{table:1},
which indicates that $T_{IS}^{-1}$ is almost independent of $\tau_L$.
i.e., $T_{IS}^{-1}$ is not very sensitive to the distance 
from the photo-excited electrons, at least, up to about 50 nm.

\begin{table}[t]
\caption{Photon-energy ($E_p$) and IR-irradiation-time ($\tau_L$) 
dependences of the cross-relaxation rates ($T_{IS}^{-1}$) measured at 10 K and 6.346 T.}
\label{table:1}
\begin{tabular}{@{\hspace{\tabcolsep}\extracolsep{\fill}}ccc} \hline \hline
$E_p$ (eV)& $\tau_L$ (s) & \hspace{5mm} $T_{IS}^{-1}$ ($10^3$ s$^{-1}$) \hspace{5mm} \\ \hline

1.386     & 60  & $8.8 \pm 0.8$ \\
1.407     & 60  & $9.4 \pm 0.8$ \\
          & 240 & $10.8 \pm 2.0$ \\
\hline \hline
\end{tabular}
\end{table} 


The cross relaxation rate $T_{IS}^{-1}$ 
can be calculated provided that all the internuclear couplings are given.
In the following, we calculate $T_{IS}^{-1}$
using the formalism by Demco et al.\cite{demco75,mehring83}
and compare the results with the experimental values.
We show
that the contribution of nuclear dipolar couplings to $T_{IS}^{-1}$
is too small to account for the experimental values,
and that indirect $J$-couplings are inevitable to account for it.

\begin{figure}[b]
 \begin{center}
   \includegraphics[scale=0.6]{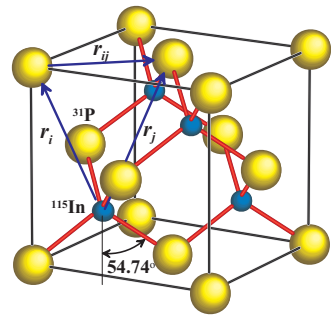}
 \end{center}
\caption{\label{crystal structure}
(Color online) Crystal structure of InP.}
\end{figure}

The cross polarization process with spin-locking can be expressed
in terms of the $x$-component of the cross polarization spectrum density $J_x(\omega)$.
Assuming a Gaussian function for $J_x(\omega)$,
$T_{IS}^{-1}$ is approximated by,\cite{demco75,mehring83}
\begin{equation}
 \frac{1}{T_{IS}}=\frac{\sqrt{\pi}}{4} \sin^2\theta_S \sin^2\theta_I  M_{2}^{IS}\tau_c 
\exp (-\Delta\omega_e^2\tau_c^2/4).
\label{ITIS}
\end{equation}
Here, $\tau_c$ is the correlation time for $J_x(\omega)$ given by,
\begin{equation}
  \frac{1}{\tau_c^2}=\frac{1}{6}P_2(\cos\theta_I)^2I(I+1)\frac{5S_4+18S_3}{S_1},
\label{tauc}
\end{equation}
where $P_2(x)$ is the second Legendre polynomial, 
and three $S_i$ values are given by the following lattice-sums;\cite{S2}
\begin{eqnarray}
  S_1&=&\sum_i B_i^2 \nonumber\\
  S_3&=&\sum_{i \neq j} A_{ij}^2 B_i B_j \nonumber\\
  S_4&=&\sum_{i \neq j} A_{ij}^2 (B_i-B_j)^2.
\label{S}
\end{eqnarray}
$A_{ij}$ and $B_i$ are the coefficients of the secular terms 
in the homo ($I_i-I_j$) and hetero ($S-I_i$) nuclear couplings, respectively.
If only the nuclear dipolar couplings are responsible for them,
they are given by,
\begin{eqnarray}
  A_{ij} \equiv -\gamma_I^2 \hbar P_2(\cos\theta_{ij})/r_{ij}^3
 =\gamma_I^2 \hbar \frac{1-3 \cos^2 \theta_{ij}}{2r_{ij}^3}, \nonumber\\
  B_i \equiv  -2\gamma_I \gamma_S \hbar P_2(\cos\theta_{i})/r_i^3
 =\gamma_I \gamma_S \hbar \frac{1-3 \cos^2 \theta_i}{r_i^3}.
\label{AB}
\end{eqnarray}
Here, $\gamma_I$ and $\gamma_S$ are the respective gyromagnetic ratios.
$\textbf{r}_{ij}$ and $\textbf{r}_{i}$ are the vectors 
corresponding to the $I_i-I_j$ and $S-I_j$ bondings,
and $\theta_i$ and $\theta_{ij}$ are the angles between $\textbf{H}_0$
and the corresponding vectors, respectively (see Fig. \ref{crystal structure}).
$M_2^{IS}$ is the second moment for the heteronuclear couplings,
which is given in the case of the dipolar couplings by,
\begin{equation}
 M_2^{IS,d}=\frac{1}{3}I(I+1)S_1.
\label{M2}
\end{equation}

\begin{figure}[b]
 \begin{center}
   \includegraphics[scale=0.5]{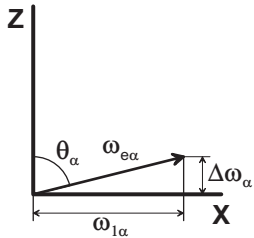}
 \end{center}
\caption{\label{off-resonance factors}
Definitions of the off-resonance factors $\omega_{e\alpha}$, 
$\omega_{1\alpha}$, $\Delta\omega_{\alpha}$ and $\theta_{\alpha}$,
where $\alpha$ corresponds to $I$ or $S$.
X and Z are the coordinates in the rotating frame.
}
\end{figure}

The off-resonance factors $\Delta\omega_e$, $\theta_I$ and $\theta_S$ are 
schematically shown in Fig. \ref{off-resonance factors}.
They are defined as,
\begin{eqnarray}
 \Delta\omega_e &\equiv &\omega_{eS}-\omega_{eI},\nonumber\\
 \theta_{I} &\equiv & \tan^{-1}(\omega_{1I}/\Delta\omega_{I}),\nonumber\\
 \theta_{S} &\equiv & \tan^{-1}(\omega_{1S}/\Delta\omega_{S}),
\end{eqnarray}
where $\omega_{1I}$ and $\omega_{1S}$ are the pulse intensities ($H_1$) 
in units of angular frequency, 
and $\Delta\omega_{\alpha}$ and $\omega_{e\alpha}$ ($\alpha=I, S$) are 
the offsets and the effective $H_1$ fields defined by,
\begin{eqnarray}
 \Delta\omega_{\alpha} &\equiv & \omega_{0\alpha}-\omega_\alpha,\nonumber\\ 
 \omega_{e\alpha}^2    &\equiv & \Delta\omega_{\alpha}^2+\omega_{1\alpha}^2,
\end{eqnarray} 
where $\omega_{0\alpha}$ and $\omega_{\alpha}$ 
are the resonance and the transmitter angular frequencies, respectively.
If both $I$ and $S$ are on-resonances ($\Delta\omega_{\alpha}=0$) 
and the Hartmann-Hahn condition is fulfilled ($\omega_{1I}=\omega_{1S}$),
\begin{eqnarray}
\Delta\omega_e &= 0,\nonumber\\
\theta_I, \theta_S & = \pi/2, 
\label{hh condition}
\end{eqnarray}
so that Eq. (\ref{ITIS}) is reduced to,
\begin{equation}
 \frac{1}{T_{IS}}=\frac{\sqrt{\pi}}{4}M_2^{IS}\tau_c.
\label{TIS simple}
\end{equation}

Now, let us calculate the dipolar contributions to $T_{IS}^{-1}$ in our case.
The off-resonance effect is negligible in our experiments because,
\begin{eqnarray}
 \Delta \omega_I/2\pi & =  -1.2 & [\textnormal{kHz}],\nonumber\\
 \Delta \omega_S/2\pi & =   1.0 & [\textnormal{kHz}],\nonumber\\
 \omega_1/2\pi        & =  16.7 & [\textnormal{kHz}]
\end{eqnarray}
so that,
$\omega_{eI} \approx \omega_{eS}$, and $\sin\theta_I =\sin\theta_S= 0.998$,
i.e., the conditions in Eq. (\ref{hh condition}) are nearly fulfilled.

The estimation of $\tau_c$ requires the calculation of lattice sums in Eq. (\ref{S}). 
In the present case,
all the nearest neighbor $I_i$ spins are at the magic angle positions 
($\theta_{i} = 54.74 ^{\circ}$) as seen in Fig. \ref{crystal structure},
so that the summations in Eq. (\ref{S}) start with the second nearest neighbor sites.
Using the lattice constant $a=5.87 {\rm \AA},$\cite{madelung04} 
the $S$-values in Eq. (\ref{S}) are calculated as, 
\begin{eqnarray}
 S_1 & = 5.50  \times 10^6    & [\textnormal{rad}^2/\textnormal{s}^2],\nonumber\\ 
 S_3 & = 4.67 \times 10^{12} & [\textnormal{rad}^4/\textnormal{s}^4],\nonumber\\
 S_4 & = 2.59 \times 10^{13} & [\textnormal{rad}^4/\textnormal{s}^4].
\label{S dipolar}
\end{eqnarray}
Equations (\ref{tauc}) and (\ref{M2}) along with $S_i$ ($i$=1, 2, 4) 
in Eq. (\ref{S dipolar}) yield 
$\tau_c= 9.08 \times 10^{-4}$ s and $M_2^{IS,d} = 1.37 \times 10^6$ rad$^2$/s$^2$.
Inserting these values into Eq. (\ref{TIS simple}),
one finally obtains,
\begin{equation}
1/T_{IS}^d = 5.5 \times 10^2 \hspace{0.2cm}[\textnormal{s}^{-1}].
\label{tISd}
\end{equation}
This value is by one order of magnitude smaller 
than the experimental values shown in Table \ref{table:1}.
This difference is presumably due to the $J$-couplings
with the nearest-neighbor $^{31}$P nuclei, $J_{IS}$.

The inclusion of $J_{IS}$ into the calculation causes 
changes in both $\tau_c$ and $M_2^{IS}$.
The correlation time $\tau_c$ in the presence of $J_{IS}$ is given 
by Eqs. (\ref{tauc}) and (\ref{AB}), 
but $2 \pi J_{IS}$ should be added to $B_i$ for the four nearest neighbor $^{31}$P.
\begin{equation}
  B_i =  -2\gamma_I \gamma_S \hbar P_2(\cos\theta_{i})/r_i^3 + 2 \pi J_{IS}.
\end{equation}
Here, we neglect the small next-nearest-neighbor homonuclear $J$-couplings 
$J(^{31}$P-$^{31}$P) of the order of 10 Hz.\cite{iijima03}
On the other hand, $M_2^{IS}$ is given by,
\begin{eqnarray}
 M_2^{IS}&=&M_2^{IS,d}+M_2^{IS,J},\label{M2-total}\\
 M_2^{IS,J}&=&\frac{1}{3}I(I+1) \sum_{i=nn}^4 (2\pi J_{IS})^2,
\end{eqnarray}
where $M_2^{IS,J}$ is the contribution from the $J$-couplings.
Note that no cross-terms between the dipolar and $J$-couplings exist 
in Eq. (\ref{M2-total}) because of the absence of the dipolar couplings 
between the nearest-neighbor sites.\cite{tomaselli98}
The value of $J_{IS}$ is determined so that the observed $T_{IS}^{-1}
= (0.9 \pm 0.3) \times 10^4$ s$^{-1}$ in Table \ref{table:1} is reproduced.
As a result, we found that it falls within the range $|J_{IS}|=2.3 \pm 0.5$ kHz.
We assume rather large error in $T_{IS}^{-1}$
taking into account the fact 
that $J_x(\omega)$ is approximated by Gaussian for simplicity.
\cite{demco75,mehring83}

The presence of $J_{IS}$ is consistent with the line width
obtained experimentally.
Assuming the Gaussian form of the spectrum,
the line width $\Delta \nu_{1/2}$ is given with the second moment $M_2$ by,\cite{mehring83}
\begin{equation}
 \Delta \nu_{1/2} =\frac{2\sqrt{2\ln2 \cdot M_2}}{2\pi}.
\end{equation}
Provided that $J_{IS}=0$, 
$M_2$ is given by the lattice sums of the homo- and heteronuclear 
dipolar couplings.
In the present case, it is calculated as, \cite{abragam61,hashi02}
\begin{eqnarray}
 M_2^d & = & M_2^{SS,d}+M_2^{IS,d} \nonumber\\
       & = & 4.1 \times 10^7 [\textnormal{rad}^2/\textnormal{s}^2],
\end{eqnarray}
which yields $\Delta \nu_{1/2}^d$ = 2.4 kHz.
In reality, it is found that 
the $^{115}$In spectrum is rather close to Lorentzian
with the full width at half of the maximum intensity (FWHM) 
of $4.0 \pm 0.2$ kHz.\cite{goto07a}
The observed FWHM is 1.7 times as large as $\Delta \nu_{1/2}^d$.
Assuming that $J_{IS}=1.4$ kHz, $M_2$ is calculated as,
\begin{eqnarray}
 M_2 & = & M_2^{SS,d}+M_2^{IS,d}+ M_2^{IS,J}\nonumber\\
     & = & 1.2 \times 10^8 [\textnormal{rad}^2/\textnormal{s}^2],
\end{eqnarray}
which yields $\Delta \nu_{1/2}$ = 4.1 kHz, reproducing the experimentally observed FWHM.
The assumed $J_{IS}$ =1.4 kHz is rather small compared 
to that estimated from $T_{IS}^{-1}$, $|J_{IS}| \approx 2.3 \pm 0.5$ kHz.
This is probably due to the ambiguities in the estimation of $T_{IS}^{-1}$
as well as the determination of the FWHM in the spectrum
whose shape is not a Gaussian.


Tomaselli et al. discussed the $J$-couplings in InP
in the cross polarization/magic angle spinning (CP/MAS) experiments.\cite{tomaselli98} 
Assuming $J_{\textnormal{aniso}}$ of the pseudo-dipolar type,
\begin{equation}
 J(\theta)=J_{\textnormal{iso}}+2J_{\textnormal{aniso}}P_2(\cos\theta),
\label{pseudo-dipolar}
\end{equation}
with $\theta$ being the angle between the nearest-neighbor 
$^{31}$P-$^{115}$In bond and the magnetic field,
they determined the isotropic and anisotropic parts of the $J$-coupling 
as $|J_{\textnormal{iso}}|=225 \pm 10$ Hz
and $|J_{\textnormal{aniso}}|=(813 \pm 50$) or ($1733 \pm 50$) Hz.
A similar conclusion has been reported by Iijima et al.\cite{iijima03} 
In the present case where $\theta=54.74^{\circ}$,
the anisotropic part of Eq. (\ref{pseudo-dipolar}) is zero,
so that $J_{IS}=J_{\textnormal{iso}}$,
whereas the value $|J_{IS}| \approx 2.3$ kHz obtained in the present study
is much greater than $|J_{\textnormal{iso}}| \approx 0.23$ kHz.

One possible explanation for the large $J_{IS}$ at $\theta=54.74^{\circ}$
is that the angular dependence of $J$ 
is not of a simple pseudo-dipolar type, 
but of the anisotropic pseudo-exchange type,\cite{bloembergen55}
\begin{equation}
 J(\theta)=J_{||}\cos^2\theta+J_{\perp}\sin^2\theta,
\end{equation}
which is a generic form of the angular dependent $J$-coupling
including the pseudo-dipolar one as a special case.
In this case, $J(54.74^{\circ})=(J_{||}+2J_{\perp})/3$,
which yields non-zero value except for $J_{||}=-2J_{\perp}$
corresponding to the pseudo-dipolar case.
The present data are still insufficient 
to determine the angular dependence of the anisotropic $J$.
Nevertheless, 
they show that the $J$-coupling is not of a simple pseudo-dipolar type.
Since the dipolar type angular dependence is averaged out in the MAS experiments,
the determination of the angular dependence of $J$ may require 
measurements of cross relaxation times 
in static cross polarization experiments.

\section{Buildup time in the optical pumping process\label{buildup time}}

In the previous section, 
we have shown that the characteristic time 
for the cross polarization process
provides us with information on the heteronuclear couplings
responsible for the polarization transfer.
A similar argument is possible for the characteristic time 
in the nuclear polarization process by the optical pumping, 
i.e., the buildup time $T_\textnormal{b}$.
It provides us with information on the hyperfine couplings
responsible for the nuclear spin orientation.

There are two types of possible hyperfine interactions in semiconductors, 
i.e., Fermi contact and dipolar interactions.
In the former case, 
photo-excited electrons are captured at shallow donor levels
whose wave functions have the diameter of the order of 100 \AA.
The nuclear spins inside the wave functions are directly polarized
through the flip-flop terms in the Fermi contact interaction ($I^+S^-+I^-S^+$).\cite{S}
In the latter case, on the one hand, 
photo-excited electrons are localized at donor sites such as deep centers.
Since little Fermi interactions exist with the surrounding nuclei,
the nuclear spins near the donor sites are polarized
through the non-secular terms of the dipole interaction ($I^+S^z+I^-S^z$).
That is, the type of hyperfine coupling is closely related to 
the state of the polarized electrons,
so that its elucidation is essential 
to understand the mechanism of the optical nuclear orientation.

Many authors have addressed this issues so far.
In the 1970's, Bagraev et al. examined the buildup time of $^{29}$Si 
in the presence of deep centers in compensated silicon
and argued the types of hyperfine couplings responsible for the optical orientation
in this material.\cite{bagraev77}
More recently, Patel et al. addressed this issue in InP.\cite{patel99}
He proposed that the two mechanisms can be distinguished 
from the difference in the sign of the nuclear polarization 
relative to that at thermal equilibrium, 
and concluded that it is the dipolar coupling
that causes nuclear polarization in undoped n-type InP.
Paravastu et al. suggested in the case of semi-insulating GaAs, 
that the photo-excited electrons localized at donor sites
cannot be solely responsible for macroscopic nuclear polarization.\cite{paravastu04} 
A factor that brings complications into the arguments
is the presence of nuclear spin diffusion,
which is believed to convey polarization farther away
from the photo-excited electrons to achieve bulk nuclear polarization. 
Goehring et al. pointed out in InP nanoparticles that 
the nuclear spin diffusion process is rather slow,\cite{goehring03}
suggesting that the spin diffusion might be rather insufficient 
to convey nuclear polarization in bulk materials.


Here, we show that the nuclear-site dependence of $T_\textnormal{b}$ 
provides a clue to identify hyperfine couplings responsible for the buildup.
We previously reported the nuclear-site dependence of 
$T_\textnormal{b}^{-1}$ in InP:Fe,\cite{goto04b}
which is summarized in Table \ref{table:2},
together with that of the spin-lattice relaxation rate $T_1^{-1}$ at 300 K.
\cite{Ep-dependence}
At first sight, it seems rather peculiar
that the values of $T_\textnormal{b}^{-1}$ at $^{115}$In and $^{31}$P 
are of the same order,
although those of $T_1^{-1}$ are different 
by four orders of magnitude between them.
If the Fermi contact were responsible 
for both $T_1^{-1}$ and $T_\textnormal{b}^{-1}$,
$T_\textnormal{b}^{-1}$ at $^{115}$In would be forty five times 
as long as that at $^{31}$P.\cite{goto04b}
This result indicates that $T_\textnormal{b}^{-1}$ and $T_1^{-1}$ 
are subject to different mechanisms from each other.
In fact, we show in the following 
that $T_1^{-1}$ is primarily caused by the Fermi contact interaction
with conduction electrons,
while $T_\textnormal{b}^{-1}$ is mainly caused by the dipolar interactions
with localized electrons.
The nuclear-site dependence of $T_1^{-1}$ stems from
the difference in the probability of electrons at the nuclear sites,
while that of $T_\textnormal{b}^{-1}$ stems from
the difference in the distance from the localized electrons.

\begin{table}[t]
\caption{The nuclear-site dependences of 
the buildup rate ($T_\textnormal{b}^{-1}$)
with the photons of $E_p$ = 1.420 eV and $\sigma^{+}$ at 4.2 K
and the spin-lattice relaxation rate ($T_1^{-1}$) at 300 K.\cite{goto04b,Ep-dependence}} 
\label{table:2}
\begin{tabular}{@{\hspace{\tabcolsep}\extracolsep{\fill}}ccc} \hline \hline
 & $T_\textnormal{b}^{-1}$ (10$^{-4}$ s$^{-1}$) \hspace{4mm} & $T_1^{-1}$ (s$^{-1}$)\\
\hline
$^{115}$In \hspace{2mm}& $ 6.3 \pm 0.2 $ \hspace{4mm} & $(1.0 \pm 0.1) \times 10^{1}$\\ 
$^{31}$P   \hspace{2mm}& $ 3.4 \pm 0.2 $ \hspace{4mm} & $(2.0 \pm 0.1) \times 10^{-3}$\\ 
\hline \hline
\end{tabular}
\end{table}

For the Fermi contact interaction with thermally excited electrons,
$T_1^{-1}$ is given by,\cite{abragam61}
\begin{equation}
 \label{T1 by Fermi couplings}
 \frac{1}{T_1}=\frac{64}{9}\pi N \eta^2\gamma_e^2 \gamma_n^2 
( \frac{m^3 k_{\textnormal B} T}{2\pi} )^{1/2},
\end{equation}
with $\eta$ being the probability of electrons/holes at the nuclear site,
and $N$ being the carrier density.
Hence, the large difference in $T_1^{-1}$ between $^{31}$P and $^{115}$In
originates from that in $\eta$.
At indium sites, 
the conduction band consists mainly of s-orbitals and has large $\eta$,
while it is small at phosphor sites
where the wave function mainly consists of p-orbitals of the valence band.

The Fermi contact interaction, however, is less effective 
for non-degenerated trapped electrons because of the following reason.
To conserve energy in the flip-flop process ($I^+S^-+I^-S^+$),
the electrons should be excited to the state with the small excitation energy 
of $\hbar\omega_{0I}$ corresponding to the nuclear Zeeman energy,
while no such excited states are available at Fermi level 
in the non-degenerated electrons.
The dipolar interaction, on the other hand,
contains non-secular terms such as $(I^+S^z+I^-S^z)$,
which flip nuclear spins $I$ without flipping electron spins $S$.
In this case, $T_\textnormal{b}^{-1}$ is given by,\cite{abragam61,patel99} 
\begin{eqnarray}
 \label{T1 by dipolar couplings}
 \frac{1}{T_\textnormal{b}} & = & \frac{3}{2}S(S+1)J^1(\omega_I)\nonumber\\
 & = & \frac{2}{5}\gamma_S^2\gamma_I^2\hbar^2\langle r^{-6} \rangle 
 S(S+1) \frac{\tau_s}{1+\omega_I^2\tau_s^2},
\end{eqnarray}
which contains no $\eta$.
Hence, contrary to the case of the Fermi contact interactions
where $\eta$ is the origin of the nuclear-site dependence, 
the nuclear-site dependence in this case stems from the differences 
in the gyromagnetic ratio ($\gamma_I$) and 
the lattice-averaged $r^{-6}$, i.e., $\langle r^{-6} \rangle$.

The nuclear-site dependence of $T_\textnormal{b}^{-1}$ allows us to estimate
the ratio of $\langle r^{-6} \rangle$ between $^{31}$P and $^{115}$In.
Taking into account the fact that $\omega_I^2\tau_s^2 \ll 1$,
Eq. (\ref{T1 by dipolar couplings}) yields,
\begin{equation}
 \langle r^{-6} \rangle \propto \frac{T_\textnormal{b}^{-1}}{\gamma_I^2}.
\end{equation}
Using the relation $(^{31}\gamma/^{115}\gamma)^2=3.41$
and $T_\textnormal{b}^{-1}$ listed in Table \ref{table:2},
one obtains, 
\begin{equation}
 \label{distance}
 \frac{^{115}\langle r^{-6} \rangle}{^{31} \langle r^{-6} \rangle} \approx 
 \left ( \frac{^{31}\gamma}{^{115}\gamma} \right ) ^2 \times
 \frac{(^{115}T_\textnormal{b})^{-1}}{(^{31}T_\textnormal{b})^{-1}} 
 = 6.3,
\end{equation}
which means that 
indium nuclei are closer to the polarized electrons in average.
This result is consistent with the values estimated from the lattice sums of $r^{-6}$
as shown below.
The calculation of the lattice sums of $r^{-6}$ can be performed
with the same calculation scheme 
as that used in Eq. (\ref{S dipolar}) in \S \ref{cross relaxation}.
Assuming that the polarized electrons are localized at phosphor sites,
the calculation yields,
\begin{equation} 
 \label{distance-lattice-sum}
 \frac{^{115}\langle r^{-6} \rangle}{^{31} \langle r^{-6} \rangle} =
\frac{\sum_i (r_i-r_0)^{-6}}{\sum_j (r_j-r_0)^{-6}} = 6.16,
\end{equation}
where $i=^{115}$In and $j=^{31}$P and 
$r_0$ is the position of the phosphor site 
at which the polarized electron is localized.
The agreement between Eq. (\ref{distance}) (experiment) and 
Eq. (\ref{distance-lattice-sum}) (calculation) is quite satisfactory. 

To summarize,
the experimental finding that $^{31}T_\textnormal{b}^{-1}$ 
and $^{115}T_\textnormal{b}^{-1}$ in InP:Fe
are of the same order indicates that 
the dipolar coupling is mainly responsible for $T_\textnormal{b}^{-1}$
in this material.
This is consistent with the conclusion by Patel et al.
deduced from the sign of the nuclear polarizations.\cite{patel99}
The present data also indicate that 
the photo-excited electrons may be located at phosphor sites,
which may be related to antisites or iron trapping centers.
This example shows that 
the nuclear-site dependence of $T_\textnormal{b}^{-1}$ provides us 
with information on the types of hyperfine couplings 
responsible for the nuclear spin orientation in the optical pumping process.

\section{Conclusion}

In conclusion, we have investigated the characteristics of 
heteronuclear and hyperfine couplings in optically oriented semiconductors 
using the semi-insulating InP:Fe.
We have focused on the time scales in the polarization transfer processes, i.e.,
the cross-relaxation time $T_{IS}$ in the cross polarization 
and the buildup time $T_\textnormal{b}$ in the optical nuclear orientation.
We find that $T_{IS}^{-1}$ is greater than
that expected from the nuclear dipolar couplings by one order of magnitude.
This discrepancy can be accounted for by 
assuming the $J$-couplings of the order of 2 kHz
between nearest-neighbor $^{115}$In-$^{31}$P.
The angular dependence of the $J$-coupling is inconsistent 
with that of a simple pseudo-dipolar type previously assumed, 
suggesting different anisotropy of the $J$-coupling.
On the other hand,
we show that the nuclear-site dependence of $T_\textnormal{b}^{-1}$ 
provides a clue to identify the hyperfine coupling 
responsible for the optical nuclear orientation.
We find in the case of InP:Fe
that $^{31}T_\textnormal{b}^{-1}$ and $^{115}T_\textnormal{b}^{-1}$ 
are of the same order,
indicating that the electron-nuclear dipolar coupling 
is primarily responsible for the nuclear spin orientation.
It is also suggested that the photo-excited electrons 
are likely to be located at phosphor sites.

\section*{ACKNOWLEDGMENTS}

This work was partially supported 
by Special Coordination Funds for Promoting Science and Technology 
of the Ministry of Education, Culture, Sports, Science and Technology (MEXT) of Japan,
Industrial Technology Research Grant Program 
from New Energy and Industrial Technology Development Organization (NEDO) of Japan, 
and Grant-in-Aid for Basic Research 
from Japan Society for Promotion of Science (JSPS).


\begin{thebibliography}{99}
\bibitem{S2} The second $S$-value $S_2$, also defined in Refs. \cite{demco75,mehring83}, is not shown here because it is irrelevant to the present calculations.
\bibitem{S} Note that $S$ in this case represents electron spins 
 and not the secondary nuclear spins in \S \ref{cross relaxation}.
\bibitem{Ep-dependence} It was confirmed that $T_\textnormal{b}^{-1}$ is almost independent 
 of $E_p$ and $\sigma^{\pm}$.\cite{goto04b}
\end{thebibliography}

\end{document}